\documentclass{jjap3}
\usepackage{txfonts}
\usepackage{graphicx}
\usepackage{epstopdf}
\usepackage{color}

\title{
Electron-transport properties of ethyne-bridged diphenyl zinc-porphyrin molecules
}

\author{
Huy Duy Nguyen$^{1}$\thanks{E-mail address: huy@cp.prec.eng.osaka-u.ac.jp} and Tomoya Ono$^{2,3}$
}
\inst{
$^{1}$ Graduate School of Engineering, Osaka University, Suita, Osaka 565-0871, Japan \\
$^{2}$ Center for Computational Sciences, University of Tsukuba, Tsukuba, Ibaraki 305-8577, Japan \\
$^{3}$ JST-PRESTO, Kawaguchi, Saitama 332-0012, Japan
}

\abst{
We investigate the electron-transport properties of ethyne-bridged diphenyl zinc-porphyrin molecules suspended between gold (111) electrodes by first-principles calculations within the framework of density functional theory. It is found that the conductance of a molecular junction in which phenyl and porphyrin rings are perpendicular is reduced by three orders of magnitude compared with that of a junction in which the phenyl and porphyrin rings are coplanar. In the coplanar configuration, electrons are transmitted through $\pi$ states, which extend over the whole molecule. In the perpendicular configuration, the conductance is suppressed because of the reduction of electron hopping between $\pi$ states of the phenyl ring and $\sigma$ states of the porphyrin ring.
}

\begin{document}
\maketitle

\section{\label{sec1}Introduction}

Controlling the electron transport through molecules that bridge two electrodes is a basic step toward the development of sophisticated nanoscale devices \cite{Nitzan,Xu}. In recent years, the results of experimental \cite{Joachim,Tans,Moresco} and theoretical studies \cite{Xue,Ventra,Orellana,Emberly} have predicted a promising future for nanodevices utilizing molecular junctions. Among many of the applications, several schemes have been proposed for the design and construction of molecular switches \cite{Moresco,Orellana,Emberly}. The key idea is to find two or more distinct states of molecules with significantly different conductance by applying an external bias or using a scanning tunneling microscope (STM) tip to manipulate the system \cite{Ghosh,Moresco}. In addition, with the advance of laser techniques, intense laser pulses can be used to manipulate the torsional motion of some conjugated molecules, which may produce ultrafast molecular switches \cite{Ramakrishna,Madsen1,Madsen2}.

Among the classes of conjugated structures in which single molecule charge transport measurements have been performed, porphyrin-based molecular structures show exceptional electronic structural characteristics. By using the STM break junction method, Li et al. fabricated a molecular junction formed by ethyne-bridged diphenyl zinc-porphyrin (edzp) molecules suspended between a gold STM tip and a gold (111) substrate \cite{Li}. These systems are of interest since it has been reported that the relative angle between the phenyl and porphyrin rings could substantially affect the conductance of the molecular junctions \cite{An}. Previous first-principles calculations on ethyne-bridged diphenyl porphyrin conjugates demonstrated that the coplanar conformation of phenyl and porphyrin rings resulted in a much larger current than the perpendicular conformation, and a much current ratio of the on/off states was observed \cite{An}. However, the authors focused on free-base porphyrins, whereas in actual experiments, zinc-porphyrins are mainly employed as building blocks to tailar the properties of materials\cite{Li,Lin,Banerjee,Sedghi1,Sedghi2,Winters}. In addition, since the supercell sizes employed in their calculations were small and the molecular geometries were not optimized when the materials were sandwiched between gold (111) electrodes, a more accurate study needs to be carried out. To the best of our knowledge, a comparison of the transport properties of an edzp molecule with various torsional angles between the phenyl and porphyrin rings has not been made.

In this paper, we examine the electron-transport properties of edzp molecules suspended between gold (111) electrodes by first-principles calculations. Our results reveal that the conductance of a molecular junction in which phenyl and porphyrin rings are perpendicular is greatly suppressed compared with that of a junction in which the phenyl and porphyrin rings are coplanar. The conducting channel in the coplanar conformation is related to $\pi$ states, which extend through the whole molecule. In contrast, the transmission in the perpendicular conformation is damped because of negligile coupling between $\pi$ states of the phenyl ring and $\sigma$ states of the porphyrin ring. The rest of this paper is organized as follows. In Sect.~\ref{sec2}, details of the computational methods and the models of the molecular junctions are given. In Sect.~\ref{sec3}, results and discussion of the electron-transport properties for these systems are presented. We summarize our findings in Sect.~\ref{sec4}.

\section{\label{sec2}Computational procedures}

All calculations are performed within the framework of density functional theory using a real-space finite-difference approach \cite{Chelikowsky,Ono1,Ono2,Ono3,Ono4}, which makes it possible to determine the self-consistent electronic ground state with a high degree of accuracy by a time-saving double-grid technique \cite{Ono1,Ono2,Ono3,Ono4}. The real-space calculations eliminate the serious drawbacks of the conventional plane-wave approach, such as its inability to treat electron-transport problems accurately.

\begin{figure}[htb]
\begin{center}
\includegraphics{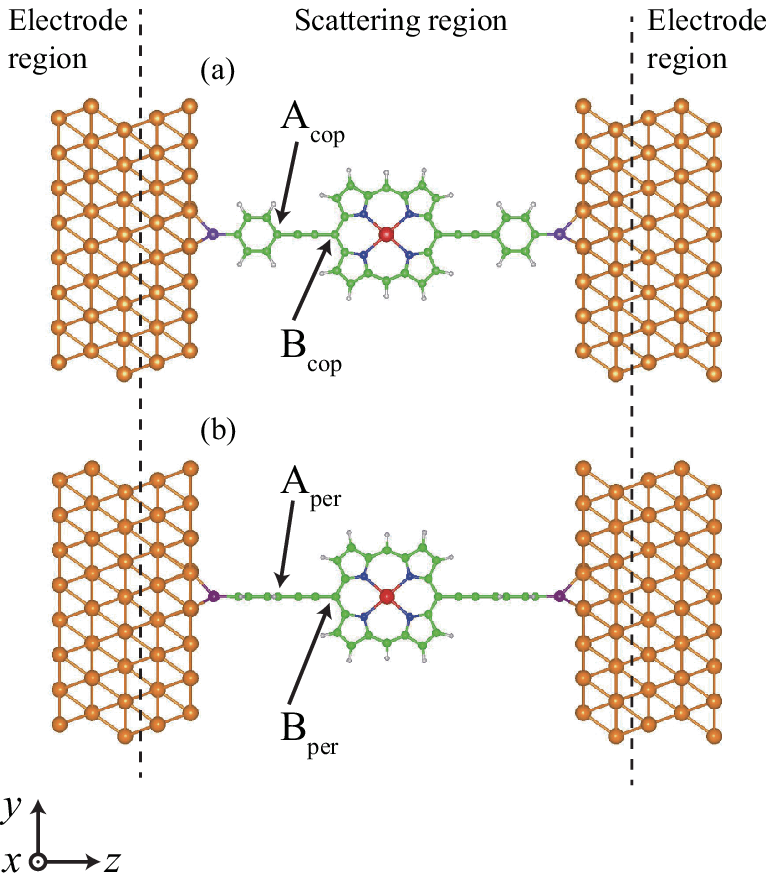}
\caption{\label{fig1} (Color) Computational models of (a) edzpc and (b) edzpp molecules. Red, green, blue, white, purple, and gold spheres indicate zinc, carbon, nitrogen, hydrogen, sulfur, and gold atoms, respectively.}
\end{center}
\end{figure}

Figure~\ref{fig1} shows the computational models of scattering regions connected to gold (111) electrodes. Such two-probe systems are divided into three parts: the left and right electrode regions, which contain three gold (111) layers with 32 atoms per each layer, and the scattering region consisting of the edzp molecule chemisorbed at the fcc sites of two gold (111) surfaces \cite{An}. We study two representative models, in which the phenyl and porphyrin rings of the edzp molecule are coplanar and perpendicular, which will be referred to as the edzpc and edzpp models, respectively. Tetragonal supercells are employed with the dimensions $L_{x}=19.95$ {\AA}, $L_{y}=11.52$ {\AA}, and $L_{z}=35.40$ {\AA} for the scattering region, where $L_{x}$ and $L_{y}$ are the lengths in the \textit{x} and \textit{y} directions parallel to the surface, respectively, and $L_{z}$ is the length in the \textit{z} direction. Structural optimizations of the scattering regions are performed by relaxing all atoms, except for those in the bottommost layers, until the remaining forces are less than 0.05 eV/{\AA}, while imposing periodic boundary conditions in all directions. The exchange-correlation interaction is treated by the local density approximation \cite{Perdew} based on density functional theory, and the projector-augmented wave method \cite{Blochl} is used to describe the electron-ion interaction.

The electron-transport properties of the edzp molecules are then computed using scattering wave functions continuing from one electrode to the other, which are obtained by the overbridging boundary-matching method \cite{Ono1,Fujimoto}. In the calculations of the scattering wave functions, the norm-conserving pseudopotentials \cite{Kobayashi} generated by the scheme proposed by Troullier and Martins \cite{Troullier} are employed instead of the projector-augmented wave method. To determine the Kohn-Sham effective potential, a supercell is used under a periodic boundary condition in all directions, and then the scattering wave functions are computed under the semi-infinite boundary condition obtained non-self-consistently. It has been reported that this procedure is just as accurate in the linear response regime but significantly more efficient than performing computations self-consistently on a scattering-wave basis \cite{Kong}. The grid spacing is set at 0.20 {\AA} and the Brillouin zone for the scattering region is sampled at the $\Gamma$ point to set up the Kohn-Sham effective potential. The conductance at zero temperature and zero bias is described by the Landauer-B\"{u}ttiker formula \cite{Buttiker}, $G = G_{0}\sum_{i,j}\lvert t_{ij}\rvert^{2}v_{i}/v_{j}$, where $t_{ij}$ is the transmission coefficient at which electrons are transmitted from an initial mode \textit{j} to a final mode \textit{i} inside the scattering region, $v_{i}$ and $v_{j}$ are the longitudinal components of the group velocity in modes \textit{i} and \textit{j}, and $G_{0} = e^{2}/h$, with \textit{e} and \textit{h} being the electron charge and Planck's constant, respectively.

\section{\label{sec3}Results and discussion}

\begin{figure}[htb]
\begin{center}
\includegraphics{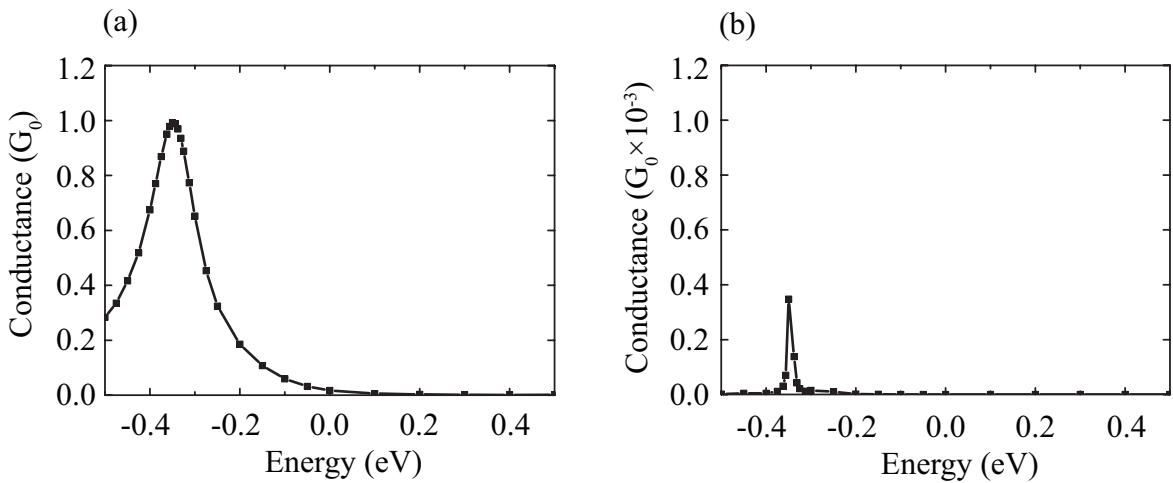}
\caption{\label{fig2} Conductances of (a) edzpc and (b) edzpp models as functions of energy of incident electrons. Zero energy is chosen to be at the Fermi level.}
\end{center}
\end{figure}

The calculated conductances of the edzpc and edzpp models, as functions of the energy of incident electrons from the left eletrode, are presented in Figs.~\ref{fig2}(a) and \ref{fig2}(b), respectively. A feature of immediate interest is that the conductance is suppressed by three orders of magnitude when the phenyl and porphyrin rings are perpendicular to each other. At the energy of $E_{F}-0.35$ eV, where $E_{F}$ is the Fermi level, the conductance of the edzpc model has one broad peak with a height of 1.00 $G_{0}$ while that of the edzpp model has one narrow peak with a height of $3.48\times10^{-3} G_{0}$.

\begin{figure}[htb]
\begin{center}
\includegraphics{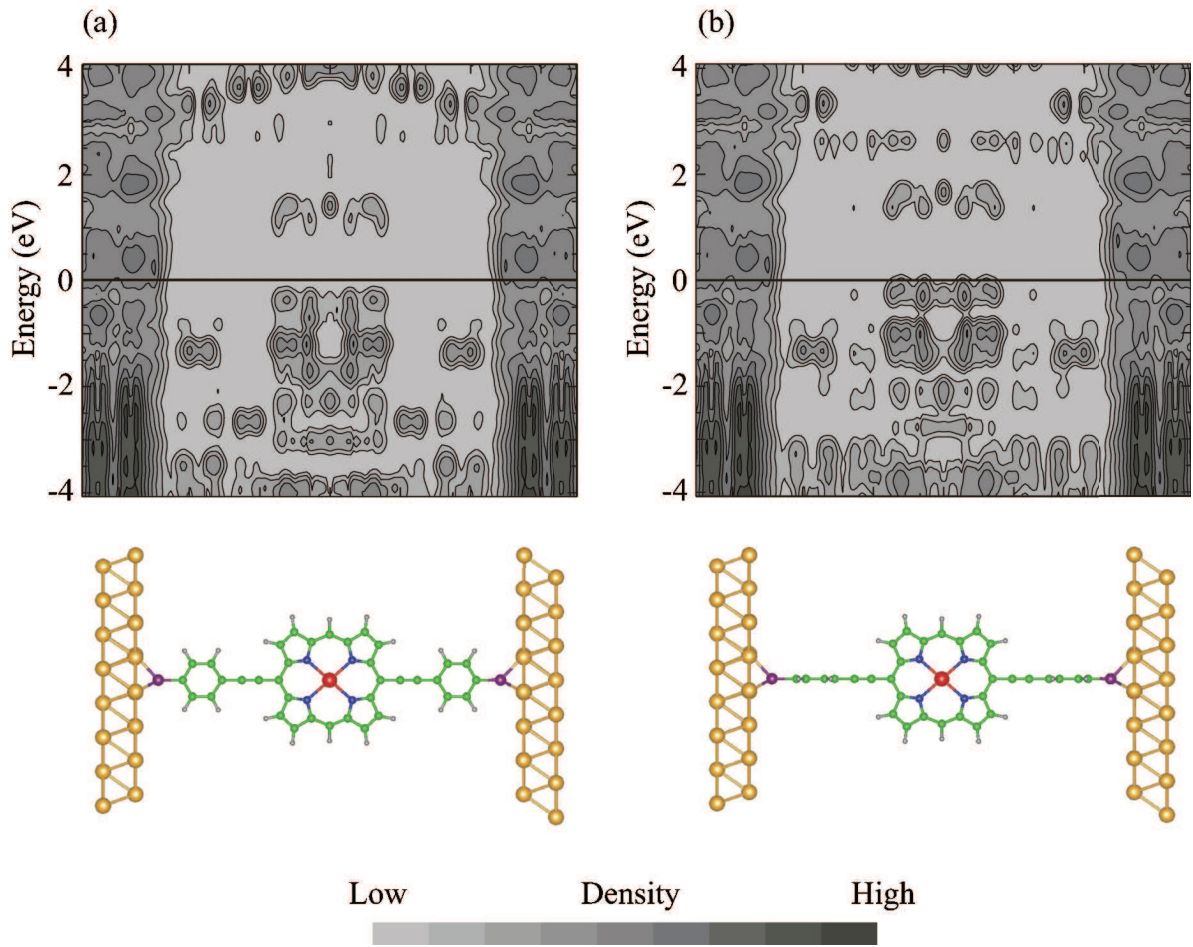}
\caption{\label{fig3} (Color) Distributions of LDOS of (a) edzpc model and (b) edzpp model integrated on plane parallel to electrode surface as functions of relative energy from the Fermi level. Zero energy is chosen to be at the Fermi level. Each contour represents twice or half the density of the adjacent contour lines, and the lowest contour is $2.78\times10^{-5}$ e/eV/{\AA}. The atomic configurations are given as a visual guide below the graph, and the symbols have the same meanings as those in Fig.~\ref{fig1}.}
\end{center}
\end{figure}

In Fig.~\ref{fig3}, we plot distribution of the local density of states (LDOS) of the two models, which are plotted by integrating them along the \textit{x}-\textit{y} plane, $\rho(z,E)=\lvert\psi(\textbf{r},E)\rvert^{2}d\textbf{r}_{\parallel}$, where $\textbf{r} = (x,y,z)$, $\psi$ is the wave function, and \textit{E} is the energy of the states. In the energy window from $E_{F}-0.50$ eV to $E_{F}$, there are electronic states localized at the porphyrin ring, which correspond to the observed peaks in the conductances of the two models. From $E_{F}$ to $E_{F}+0.50$ eV, the conductances are almost zero owing to the presence of the gap in the LDOS. 

\begin{figure}[htb]
\begin{center}
\includegraphics{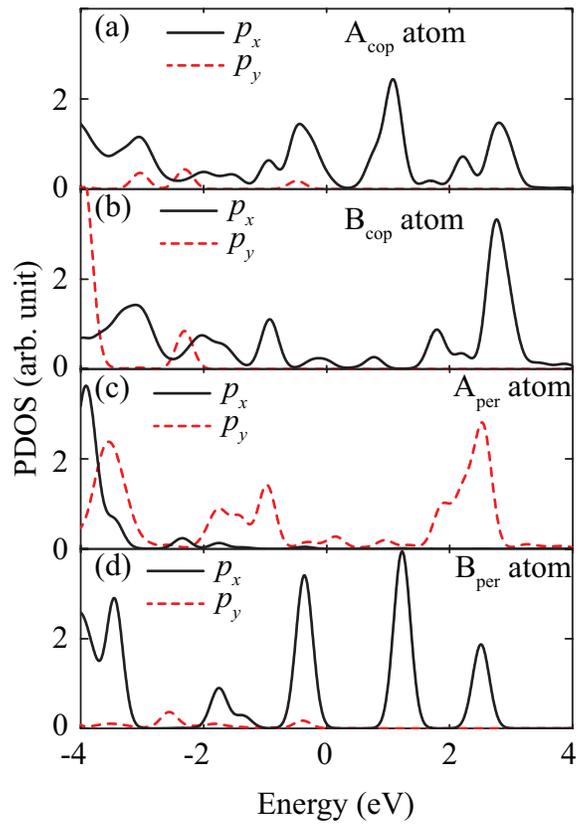}
\caption{\label{fig4} (Color online) PDOS of $p_{x}$ and $p_{y}$ orbitals of (a) $A_{cop}$ and (b) $B_{cop}$ atoms of edzpc model shown in Fig.~\ref{fig1}(a), and PDOS of (c) $A_{per}$ and (d) $B_{per}$ atoms of edzpp model shown in Fig.~\ref{fig1}(b). Zero energy is chosen to be at the Fermi level.}
\end{center}
\end{figure}

It is notable that the conductance of the edzpp model is very small although there is a large density of states at the porphyrin ring from $E_{F}-0.50$ eV to $E_{F}$. To understand the effect of the rotation of the phenyl rings on the electron-transport property, in Fig.~\ref{fig4} we plot the projected density of states (PDOS) on the $p_{x}$ and $p_{y}$ atomic orbitals of the carbon atoms indicated by the arrows in Fig.~\ref{fig1}. Around the Fermi level, $\pi$ states are expected to contribute to the electron transport \cite{An}. In the edzpc model, the electrons are propagated from the left electrode by states with the $p_{x}$ character. Since there is a substantial overlap between the PDOS on the $p_{x}$ orbital of the $A_{cop}$ atom of the phenyl ring and that of the $B_{cop}$ atom of the porphyrin ring, the conductance is large. In contrast, when the phenyl rings are rotated by $90^{\textrm{o}}$, the electrons are propagated from the left electrode by states with the $p_{y}$ character. Because the overlap between the PDOS on the $p_{y}$ orbital of the $A_{per}$ and $B_{per}$ atoms is small, the electron transport is reduced.

\begin{figure}[htb]
\begin{center}
\includegraphics{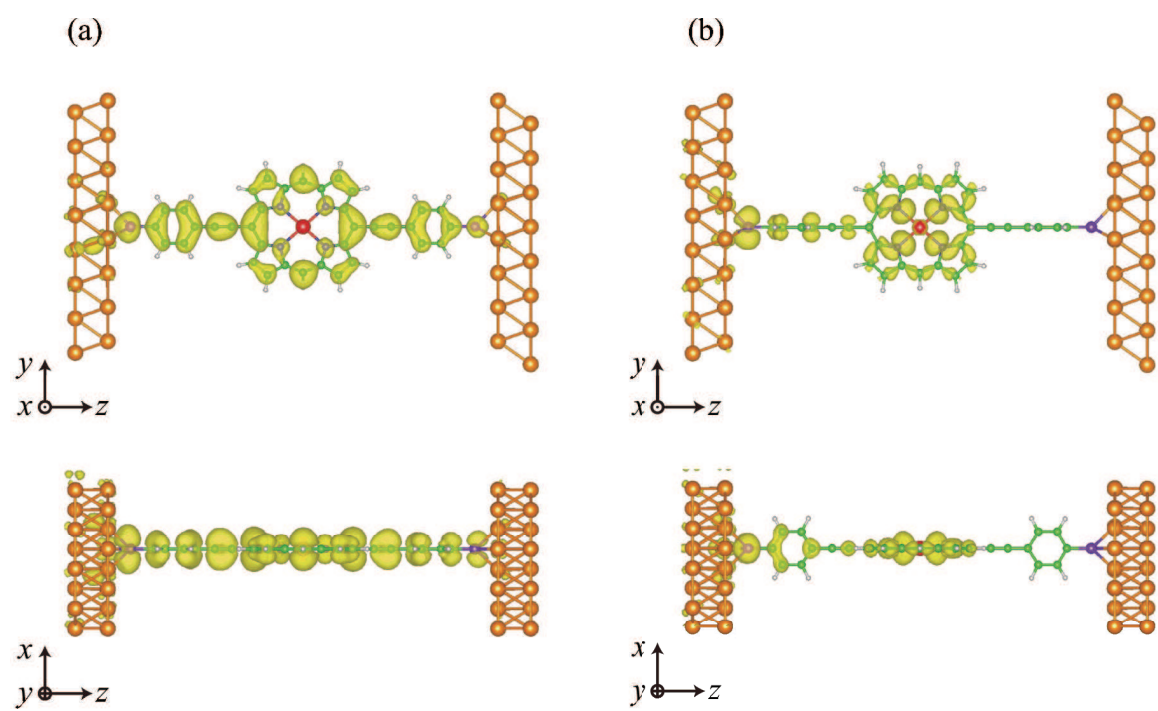}
\caption{\label{fig5} (Color) Top and side views of isosurfaces of charge density distributions of scattering waves emitted from left electrode of (a) edzpc model and (b) edzpp model at the energy of $E_{F}-0.35$ eV. The isovalue is $1.24\times10^{-3}$ \textit{e}/{\AA}$^{3}$/eV. The symbols have the same meanings as those in Fig.~\ref{fig1}.}
\end{center}
\end{figure}

Finally, to investigate which states contribute to the electron transport in more detail in Fig.~\ref{fig5}, we depict the isosurfaces of the charge density distribution of the scattering waves for electrons injected from the left electrode. The peak in the conductance of the edzpc model corresponds to the transmission of $\pi$ states of the molecule. Because of the orthorgonality between the phenyl and porphyrin rings in the edzpp model, $\pi$ states of the porphyrin ring do not conduct electrons whereas $\sigma$ states of the porphyrin ring can contribute to the electron transport. However, electrons are reflected at the junctions between the phenyl and porphyrin rings owing to the small electron-hopping probability from $\pi$ states of the phenyl ring to $\sigma$ states of the porphyrin ring, which results in the small conductance.

\section{\label{sec4}Conclusion}

We have studied the electron-transport properties of edzp molecules suspended between gold (111) electrodes by first-principles calculations. It is found that the conductance through the molecular junction is greatly suppressed when the phenyl rings are perpendicular to the porphyrin ring. The peaks of conductances of 1.00 $G_{0}$ and $3.48\times10^{-3}$ $G_{0}$ for the edzpc and edzpp models, respectively, are observed at the energy of $E_{F}-0.35$ eV. The charge density distributions of the scattering waves reveal that the conduction through the edzpc model originates from the transmission of $\pi$ states of the porphyrin ring while the conduction through the edzpp model is related to $\sigma$ states of the porphyrin ring because $\pi$ states of the porphyrin ring do not contribute to the electron transport. Our results suggest that this molecular junction can be used to construct molecular switches.

\begin{acknowledgments}
This research was partially supported by the Computational Materials Science Initiative (CMSI), the Japan Society for the Promotion of Science 'Core to Core' Program, and a Grant-in-Aid for Scientific Research on Innovative Areas (Grant No. 22104007) from the Ministry of Education, Culture, Sports, Science and Technology, Japan. The numerical calculation was carried out using the computer facilities of the Institute for Solid State Physics at the University of Tokyo and Center for Computational Sciences at University of Tsukuba.
\end{acknowledgments}

\clearpage

\end{document}